\documentclass[
amsmath,
prd,nofootinbib,floatfix,12pt,
]{revtex4}

\def\MSbar{\overline{\rm MS}}
\def\gptwo{g^{\prime 2}}
\def\lnbar{\overline{\ln}}
\newcommand\beq{\begin{eqnarray}}
\newcommand\eeq{\end{eqnarray}}
\newcommand\Tbar{\overline{T}}
\def\lsim{\mathrel{\rlap{\lower4pt\hbox{$\sim$}}
    \raise1pt\hbox{$<$}}}                
\def\gsim{\mathrel{\rlap{\lower4pt\hbox{$\sim$}}
    \raise1pt\hbox{$>$}}}            

\allowdisplaybreaks
\usepackage{graphicx}
\usepackage{setspace}
\usepackage{bm}

\begin{document}

\renewcommand{\theequation}{\arabic{section}.\arabic{equation}}
\renewcommand{\thefigure}{\arabic{section}.\arabic{figure}}
\renewcommand{\thetable}{\arabic{section}.\arabic{table}}

\title{\large \baselineskip=20pt 
Pole mass of the $W$ boson at two-loop order\\ 
in the pure $\overline{\rm MS}$ scheme
}\baselineskip=16pt 

\author{Stephen P.~Martin}
\affiliation{
{\it Department of Physics, Northern Illinois University, DeKalb IL 60115},
{\it Fermi National Accelerator Laboratory, P.O. Box 500, Batavia IL 60510}}

\begin{abstract}\normalsize\baselineskip=16pt 
I provide a calculation at full two-loop order of the complex pole 
squared mass of the $W$ boson in the Standard Model in the pure $\MSbar$ 
renormalization scheme, with Goldstone boson mass effects resummed. This 
approach is an alternative to earlier ones that use on-shell or hybrid 
renormalization schemes. The renormalization scale dependence of the 
real and imaginary parts of the resulting pole mass are studied. Both 
deviate by about $\pm 4$ MeV from their median values as the 
renormalization scale is varied from 50 GeV to 200 GeV, but the theory 
error is likely larger. A surprising feature of this scheme is that the 
2-loop QCD correction has a larger scale-dependence, but a smaller 
magnitude, than the 2-loop non-QCD correction, unless the 
renormalization scale is chosen very far from the top-quark mass.
\end{abstract}

\maketitle

\vspace{-0.35in}

\tableofcontents

\baselineskip=17pt

\section{Introduction\label{sec:intro}}
\setcounter{equation}{0}
\setcounter{figure}{0}
\setcounter{table}{0}
\setcounter{footnote}{1}

The discovery \cite{LHCdiscovery,LHCHmass} of the 125 GeV Higgs boson 
$h$ at the Large Hadron Collider (LHC) has completed the minimal 
Standard Model of electroweak symmetry breaking. Since the LHC has also 
not discovered any superpartners or other new fundamental particles, it 
is now more motivated than ever to perform precision analyses of the 
masses and interactions of the known particles of the completed theory. 
This paper concerns the complex pole mass 
\cite{Tarrach:1980up,Stuart:1991xk,Willenbrock:1991hu,Sirlin:1991fd,
Stuart:1992jf,Passera:1996nk,Kronfeld:1998di} 
of the $W$ boson,
\beq
s^W_{\rm pole}  &\equiv& M^2_W - i \Gamma_W M_W
\eeq
calculated at 2-loop order.

There have already been many studies 
\cite{Sirlin:1980nh, Marciano:1980pb, Sirlin:1983ys, Djouadi:1987gn,
Consoli:1989fg, Kniehl:1989yc, Djouadi:1993ss, Avdeev:1994db,
Chetyrkin:1995ix, Chetyrkin:1995js, Degrassi:1996mg, Degrassi:1996ps,
Degrassi:1997iy, Passera:1998uj, Jegerlehner:2001fb, Jegerlehner:2002em, Freitas:2000gg, 
Awramik:2002wn, Faisst:2003px, Awramik:2003ee, Awramik:2003rn, Schroder:2005db, 
Chetyrkin:2006bj, Boughezal:2006xk, Sirlin:2012mh, Degrassi:2014sxa, Kniehl:2015nwa}
that calculate contributions to the physical $W$ boson mass, including all 
2-loop order contributions and some QCD-enhanced effects at 3- and 
4-loop order besides. (These are reviewed in 
refs.~\cite{Sirlin:2012mh,Degrassi:2014sxa}, for example.) Indeed, the 
accuracy of the most advanced of these calculations exceeds that of the 
present paper when it comes to predicting the $W$-boson mass in terms of 
other measured quantities. However, the existing calculations have been 
done in on-shell or hybrid $\MSbar$/on-shell schemes, or use expansions 
in small squared mass ratios, as in the case of 
ref.~\cite{Jegerlehner:2001fb, Jegerlehner:2002em}. In this paper, I will provide a 
calculation that does not employ mass ratio expansions and uses a 
``pure" $\MSbar$ scheme, which means that the complete set of input 
parameters consists of only the renormalized running $\MSbar$ quantities
\beq
v,\> g,\> g',\> \lambda,\> y_t,\> g_3
\label{eq:inputs}
\eeq
at a given renormalization scale $Q$. Here, $v(Q)$ is defined to be the 
minimum of the radiatively corrected effective potential in Landau 
gauge, which is now known to full 2-loop order \cite{FJJ} with 3-loop 
contributions at leading order in $g_3$ and $y_t$ \cite{Martin:2013gka}, 
with Goldstone boson mass contributions resummed 
\cite{Martin:2014bca,Elias-Miro:2014pca}. This allows $v$ to be traded 
for the Higgs squared mass parameter $m^2(Q)$. The normalizations
of $v, m^2$ and $\lambda$ are such that the 
Higgs potential is
\beq
V &=& m^2 \Phi^\dagger \Phi + \lambda (\Phi^\dagger \Phi)^2 .
\eeq
and $\langle \Phi \rangle = v/\sqrt{2}$, with a canonically normalized 
Higgs doublet field $\Phi$. 

In principle, the input parameters should also include the other quark 
and lepton Yukawa couplings, but these make only a very small difference 
in the present paper, as discussed below. In the pure $\MSbar$ scheme 
approach, all of the complex pole masses and other observables, 
such as the Fermi decays constant, 
are outputs, to be computed in terms of the quantities in 
eq.~(\ref{eq:inputs}). In practice, global fits to data may be used 
to obtain the relationship. In this paper, the input parameters of 
eq.~(\ref{eq:inputs}) are all understood to be in the full non-decoupled 
(6-quark) Standard Model theory. Note that if the renormalization scale $Q$ is chosen between $M_W$ and $M_t$, the largest logarithms encountered in calculations of the physical masses of $W,Z,h,t$ will be at most $\ln(M_t^2/M_W^2) \approx 1.5$.

It has been argued that the experimental vector boson masses $M_{V,{\rm exp}}$ 
as measured at colliders are
related to the complex pole mass quantities by, approximately 
\cite{Bardin:1988xt}, \cite{Willenbrock:1991hu,Sirlin:1991fd}:
\beq
M_{V,{\rm exp}}^2 &=& M_V^2 + \Gamma_V^2.
\eeq
Numerically, this amounts to $M_{W,{\rm exp}} \approx M_{W} + 27$ MeV in 
the case of the $W$ boson, assuming the Standard Model prediction for 
the width. Here, $M_V^2$ is the the real part of the complex pole of the 
propagator, while $M_{V,{\rm exp}}^2$ corresponds to what is sometimes 
called the ``on-shell" mass. In the following, I will refer to $M_{W}$ 
rather than $M_{W,{\rm exp}}$. The current experimental value \cite{RPP} 
is $M_{W,{\rm exp}} = 80.385 \pm 0.015$ GeV.

At the present time, the pure $\MSbar$ scheme is not quite competitive 
in numerical accuracy with the on-shell or hybrid schemes for the 
$W$-boson mass calculation (although it is for the Higgs boson mass, 
which has been obtained to 2-loop order with the leading 3-loop 
corrections \cite{Martin:2014cxa,SMHwebpages}). However, as the 
technology for loop calculations improves, it is quite possible that 
this will change. As a matter of opinion, I find the modular approach of 
the pure $\MSbar$ scheme to be conceptually simpler, and it can be 
easily extended to include contributions from new particles beyond the 
Standard Model, and the methods used can even be applied to other vector 
bosons (such as a $W'$) in different theories. In any case, there is 
hopefully some value in being able to compare different schemes for the 
Standard Model observables, given their importance.

\section{$W$ boson complex pole mass at 2-loop order\label{sec:Mh}}
\setcounter{equation}{0}
\setcounter{figure}{0}
\setcounter{table}{0}
\setcounter{footnote}{1}

In this section, I describe the calculation of the $W$-boson complex 
pole mass. The calculation reported here is restricted to Landau gauge, 
because only in that gauge has the effective potential been evaluated to 
full 2-loop order with leading 3-loop corrections, and this is necessary 
to obtain the relationship between the Higgs vacuum expectation value 
(VEV) and the Lagrangian squared mass parameter, used implicitly in the 
calculation below. However, the complex pole mass 
\cite{Tarrach:1980up,Stuart:1991xk,Willenbrock:1991hu,Sirlin:1991fd,
Stuart:1992jf,Passera:1996nk,Kronfeld:1998di} 
is a physical observable. It is therefore independent of the gauge 
fixing parameters \cite{Gambino:1999ai}, as well as renormalization 
group invariant.

In order to obtain the $W$-boson complex pole mass, one first obtains, in 
terms of bare parameters in the regulated theory in $d=4-2\epsilon$ 
dimensions, the transverse self-energy function
\beq
\Pi(s) = \frac{1}{16 \pi^2} \Pi^{(1)}(s) 
\,+\, \frac{1}{(16 \pi^2)^2} \Pi^{(2)}(s) .
\eeq
This is obtained by constructing the $W$-boson 
self-energy function $\Pi^{WW}_{\mu\nu}(s)$
from the sum of all 1-particle-irreducible 2-point Feynman 
diagrams, and then contracting with 
$(\eta^{\mu\nu} - p^\mu p^\nu/p^2)/(d-1)$,
where $p^\mu$ is the external momentum and $s = -p^2$, using a metric with
Euclidean or ($-$,$+$,$+$,$+$) signature. 
Factors of $1/(16 \pi^2)^\ell$ are used to signify 
the loop order $\ell$. Rather than including counterterm diagrams separately, 
it is more convenient 
and efficient to do the calculation in terms of the bare 
quantities: the VEV $v_B$ and 
the bare Higgs squared mass parameter 
$m^2_B$, and the couplings
$g_B$, 
$g'_B$, and 
$\lambda_B$, 
$y_{tB}$, 
$g_{3B}$, 
and then rewrite the results in terms of the $\MSbar$ quantities. 

The finite, renormalization-group invariant, and gauge-fixing invariant complex pole squared mass can be written at 2-loop order:
\beq
s^W_{\rm pole} &=&
W_B + 
\frac{1}{16 \pi^2} \Pi^{(1)}(W_B) \,+\,
 \frac{1}{(16 \pi^2)^2} \left [\Pi^{(2)}(W_B) + 
 \Pi^{(1)\prime}(W_B) \Pi^{(1)}(W_B) \right ],
\label{eq:M2hbare}
\eeq
where $W_B = g_B^2 v_B^2/4$. The bare quantities are then eliminated in 
favor of the $\MSbar$ renormalized parameters using:
\beq
\label{eq:v2B}
v^2_B &=& \mu^{-2\epsilon} v^2 \Bigl [
1 + \frac{1}{16\pi^2} \frac{c^\phi_{1,1}}{\epsilon} 
+ \frac{1}{(16\pi^2)^2} \Bigl ( \frac{c^\phi_{2,2}}{\epsilon^2}
+ \frac{c^\phi_{2,1}}{\epsilon} \Bigr ) + \ldots \Bigr ]
,
\\
g_{B} &=& \mu^{\epsilon} \Bigl [
g + \frac{1}{16\pi^2} \frac{c^{g}_{1,1}}{\epsilon} 
+ \frac{1}{(16\pi^2)^2} \Bigl ( \frac{c^{g}_{2,2}}{\epsilon^2}
+ \frac{c^{g}_{2,1}}{\epsilon} \Bigr ) 
+ \ldots \Bigr ]
,
\\
g'_{B} &=& \mu^{\epsilon} \Bigl [
g' + \frac{1}{16\pi^2} \frac{c^{g'}_{1,1}}{\epsilon} 
+ \ldots \Bigr ]
,
\\
\lambda_B &=& \mu^{2\epsilon} \Bigl [
\lambda + \frac{1}{16\pi^2} \frac{c^\lambda_{1,1}}{\epsilon} 
 + \ldots \Bigr ]
,
\\
m^2_B &=& 
m^2 + \frac{1}{16\pi^2} \frac{c^{m^2}_{1,1}}{\epsilon} 
+ \ldots
,
\\
y_{tB} &=& \mu^{\epsilon} \Bigl [
y_t + \frac{1}{16\pi^2} \frac{c^{y_t}_{1,1}}{\epsilon} 
+ \ldots \Bigr ]
,
\\
g_{3B} &=& \mu^{\epsilon} \left [
g_3 + \ldots \right ]
\label{eq:g3B}
,
\eeq
to obtain $s^W_{\rm pole}$ in terms of the renormalized parameters. 
Here $\mu$ is the dimensional regularization scale. 
The $\MSbar$ renormalization scale $Q$ is related to it by
\beq
Q^2 = 4\pi e^{-\gamma_E} \mu^2,
\eeq
where $\gamma_E$ is the Euler-Mascheroni constant. The
counterterm coefficients were listed, in exactly 
the same conventions as in this paper,
in ref.~\cite{Martin:2014cxa}, except for:
\beq
c_{2,1}^g &=& 
\frac{35}{24} g^5 
+ 3 g^3 g_3^2 
+ \frac{3}{8} g^3 \gptwo 
- \frac{3}{8} g^3 y_t^2
,
\label{eq:cg21}
\\
c_{2,2}^g &=& \frac{361}{96} g^5
\label{eq:cg22}
.
\eeq
All of these counterterm coefficients can be obtained from the
2-loop beta functions and scalar anomalous dimension found in
refs.~\cite{MVI,MVII,Jack:1984vj,MVIII}, \cite{FJJ}; see for example
the discussion surrounding eqs.~(4.5)-(4.14) of ref.~\cite{Martin:2013gka}.

The procedure for the rest of the calculation is quite similar to that 
in ref.~\cite{Martin:2014cxa}, to which the reader is therefore referred for
some more details, in a (perhaps futile) attempt 
to avoid triggering the arXiv's self-plagiarism detector. 
The Tarasov algorithm \cite{Tarasov:1997kx} is used to reduce the
2-loop integrals to a basis set. 
The program {\tt TARCER} \cite{Mertig:1998vk} that is 
often used for this purpose
was apparently unable to handle a few of
the necessary reductions in a finite time, 
so I wrote a new Mathematica program {\tt RedTint} 
implementing the Tarasov algorithm. 
(This program will be publicly released soon.) 
After expansion in $\epsilon = (4-d)/2$, the Tarasov basis integrals
were then written in terms of a set of basis integrals
defined and described in detail in
refs.~\cite{Martin:2003qz,TSIL}. The 1-loop basis integrals are:
\beq
A(x),\>\>B(x,y)\label{eq:deAB}
\eeq
and the 2-loop basis integral list is
\beq
I(x,y,z),\>\>S(x,y,z),\>\>T(x,y,z),\>\>\Tbar(0,x,y),\>\>
U(x,y,z,u),\>\> M(x,y,z,u,v) .
\label{eq:defISTUM}
\eeq
The arguments $x,y,\ldots$ are squared masses, and  
$B, S, T, \Tbar, U, M$ also each have
an implicit dependence on the external momentum 
invariant $s = -p^2$,
while $A, B, I, S, T, \Tbar, U$ have an implicit dependence on 
the renormalization scale $Q$.
The computer program {\tt TSIL} \cite{TSIL}
can then be used for the efficient numerical
evaluation of these basis integrals. {\tt TSIL} uses Runge-Kutta
integration of differential
equations similar to that suggested in 
ref.~\cite{Caffo:1998du},
and also includes relevant
analytical results found in refs.~\cite{Martin:2003qz,
Broadhurst:1987ei, Gray:1990yh, Davydychev:1992mt, Davydychev:1993pg, 
Scharf:1993ds, Berends:1994ed, Berends:1997vk}. 

After writing bare quantities in terms of $\MSbar$ quantities and expanding
in $\epsilon$,
the tree-level squared-mass arguments of the basis integrals used 
in the final result are:
\beq
W&=& g^2 v^2/4,
\label{eq:defW}
\\
Z&=& (g^2 + g^{\prime 2}) v^2/4,
\\
t&=& y_{t}^2 v^2/2,
\\
h &=& 2 \lambda v^2
\label{eq:defh}
\eeq
and 0 for photons and gluons. As in \cite{Martin:2014cxa},
the Goldstone boson squared masses are eliminated by using
the condition for the minimization of the effective potential after resummation,
\beq
m^2 + \lambda v^2 &=&
\frac{1}{16 \pi^2} \Bigl \{
2 N_c y_t^2 A(t) 
- 3 \lambda A(h) 
- \frac{g^2}{2} [3 A(W) + 2 W]
\nonumber \\ &&
- \frac{g^2 + \gptwo}{4} [3 A(Z) + 2 Z]
\Bigr \} + \ldots,
\label{eq:resummedmincon}
\eeq
as explained in section 4 of ref.~\cite{Martin:2014bca} 
(see also \cite{Elias-Miro:2014pca} and \cite{Pilaftsis:2015cka}).
The same relation is used to eliminate $m^2$ from the 
tree-level Higgs boson squared mass, which appears as 
$h$ rather than $H = m^2 + 3 \lambda v^2$. In a future 3-loop calculation of
the $W$ (or $Z$) pole mass, 
the 2-loop version of eq.~(\ref{eq:resummedmincon}) should be used; this can be found
in eqs.~(4.18)-(4.20) of ref.~\cite{Martin:2014bca}.

The 2-loop $W$ boson squared pole mass is thus obtained, after finally 
taking $\epsilon \rightarrow 0$, as:
\beq
s^W_{\rm pole}  &=& M^2_W - i \Gamma_W M_W \>=\> 
W
+ \frac{1}{16 \pi^2} \Delta^{(1)}_W
+ \frac{1}{(16 \pi^2)^2} \left [
\Delta^{(2),{\rm QCD}}_W + \Delta^{(2),{\rm non-QCD}}_W \right ],
\phantom{xxx}
\label{eq:M2Wpole}
\eeq
where the right-hand side is a function of $v, g, g', \lambda, y_t,  g_3,Q$,
with all propagator masses expressed as $W, Z, h, t$, or $0$. 
The list of 
1-loop basis integrals used is
\beq
I^{(1)} &=& \bigl \{
A(h),\> A(t),\> A(W),\> A(Z),\> B(0, 0),\> B(0, h),\> B(0, t),
\nonumber \\ && 
B(0, Z),\> B(h, t),\> B(h, W),\> B(t, Z),\> B(W, Z)
\bigr \} ,
\label{eq:onelooplist}
\eeq
while the list of necessary 2-loop basis integrals is:
\beq
I^{(2)} &=& \bigl \{
I(0, 0, h),\> I(0, 0, t),\> I(0, 0, W),\> I(0, 0, Z),\> 
I(0, h, W),\> I(0, h, Z),\> I(0, t, W),\> 
\nonumber \\ && 
I(0, W, Z),\> 
I(h, h, h),\> I(h, t, t),\> I(h, W, W),\> I(h, Z, Z),\> 
I(t, t, Z),\> I(W, W, Z),\>
\nonumber \\ && 
S(h, h, W),\> S(h, W, Z),\> S(t, t, W),\> S(W, Z, Z),\> 
T(h, 0, 0),\> T(h, 0, t),\> 
\nonumber \\ && 
T(h, 0, W),\> T(h, W, Z),\> 
T(t, 0, 0),\> T(t, 0, h),\> T(t, 0, Z),\> T(W, 0, 0),\> 
\nonumber \\ && 
T(Z, 0, 0),\> T(Z, 0, t),\> T(Z, 0, W),\> T(Z, h, W),\> 
\Tbar(0, h, W),\> \Tbar(0, W, Z),\> 
\nonumber \\ && 
U(0, t, 0, W),\> 
U(0, t, h, t),\> 
U(0, t, t, Z),\> 
U(h, W, 0, 0),\> 
U(h, W, 0, t),\> 
\nonumber \\ && 
U(h, W, h, W),\> 
U(h, W, W, Z),\> 
U(W, 0, t, t),\> 
U(W, h, h, h),\> 
U(W, h, t, t),\> 
\nonumber \\ && 
U(W, h, W, W),\> 
U(W, h, Z, Z),\> 
U(W, Z, 0, 0),\> 
U(W, Z, h, Z),\> 
U(W, Z, t, t),\> 
\nonumber \\ && 
U(W, Z, W, W),\> 
U(Z, W, 0, 0),\> 
U(Z, W, 0, t),\> 
U(Z, W, h, W),\> 
U(Z, W, W, Z),\>
\nonumber \\ && 
M(0, 0, 0, 0, 0),\> 
M(0, 0, 0, 0, Z),\> 
M(0, 0, 0, W, 0),\> 
M(0, 0, t, t, 0),\> 
M(0, 0, t, t, Z),\> 
\nonumber \\ && 
M(0, 0, t, W, 0),\> 
M(0, t, W, 0, t),\> 
M(0, W, 0, Z, 0),\> 
M(0, W, t, h, t),\> 
\nonumber \\ && 
M(0, W, t, Z, t),\> 
M(0, W, W, 0, W),\> 
M(0, W, W, h, W),\> 
M(0, W, W, Z, W),\> 
\nonumber \\ && 
M(0, Z, t, W, 0),\> 
M(h, h, W, W, h),\> 
M(h, W, W, h, W),\> 
M(h, W, W, Z, W),\> 
\nonumber \\ && 
M(h, Z, W, W, Z),\> 
M(W, W, Z, Z, h),\> 
M(W, Z, Z, W, W)
\bigr \} .
\label{eq:twolooplist}
\eeq
In each of the $B$, $S$, $T$, $\overline T$, $U$, and $M$ integrals, 
the external momentum invariant is the tree-level squared mass, $s = W$.

The 1-loop contribution to the pole mass is:
\beq
\Delta^{(1)}_W &=&
g^2 \biggl \{ N_c |V_{tb}|^2 f(b,t,W) + 
\left [N_c (n_Q- |V_{tb}|^2) + n_L \right ] f(0,0,W)
+ \left (\frac{1}{4} - \frac{h}{12 W} \right ) A(h)
\nonumber \\ &&
+ \left (\frac{4 W}{Z} + \frac{h+Z}{12 W}  - 3\right ) A(W)
+ \left (\frac{2 W}{Z} - \frac{2}{3} - \frac{Z}{12 W} \right ) A(Z)
\nonumber \\ &&
+ \left ( \frac{4 W^2}{Z} + \frac{17 W - 4 Z}{3} - \frac{Z^2}{12 W} \right ) 
B(W,Z)
+ \left ( \frac{h}{3} - \frac{h^2}{12 W} - W  \right ) B(h,W)
\phantom{xxxx}
\nonumber \\ &&
- \frac{4 W^2}{Z} + \frac{64 W}{9} + \frac{h+Z}{6}
\biggr \}
,
\label{eq:Delta1loop}
\eeq
where 
\beq
N_c = n_Q = n_L = 3
\eeq
are the numbers of colors, quark doublets, and lepton doublets
in the Standard Model, respectively,
and the fermion-loop function is
\beq
f(x,y,s) &=& \frac{1}{6s} \Bigl \{
\bigl [(x-y)^2 + s (x+y) - 2 s^2 \bigr ] B(x,y) + 
(x-y-2s) A(x)  
\nonumber \\ && 
+ (y-x-2s) A(y)\Bigr \} + (s - 3x - 3y)/9 ,
\eeq
and the bottom quark mass and $|V_{tb}|^2$ dependence has been included.
The lighter quark and lepton 
masses can also be restored in the obvious way, by changing the 0 
arguments of the function $f$ in eq.~(\ref{eq:Delta1loop}) and introducing additional
Cabibbo-Kobayashi-Maskawa (CKM) mixing factors. 
Fortunately, however, the difference made by non-zero masses of $b,\tau,c,\ldots$ 
and the presence of CKM mixing (assuming CKM unitarity and $V_{tb} = 0.99914$ \cite{RPP}) 
is less than about 1 MeV in both $M_W$ and $\Gamma_W$ for 50 GeV $<Q<$ 200 GeV,
and is much less for $Q$ in the middle of that range, 
so those effects will be neglected for simplicity below.

Note that 1-loop contributions involving
$B(0,0)$, $B(0,Z)$ and $B(0,h)$ cancel,
when the 0 arguments correspond to Goldstone bosons and
unphysical modes of the vector bosons
in Landau gauge. This and similar cancellations
in the 2-loop order part (mentioned below) are useful checks, 
as non-cancellation of such terms would have implied imaginary 
parts of the complex pole squared mass that do not correspond 
to any real decay mode of the $W$ boson. 

The 2-loop QCD contribution is also simple enough to be written on a few lines
in terms of the basis functions:
\beq
\Delta^{(2),{\rm QCD}}_W &=& g_3^2 g^2 \Bigl (\frac{N_c^2 -1}{24} \Bigr ) \biggl [
-4 (t-W)^2 (2 + t/W) M(0,0,t,t,0) 
\nonumber \\ &&
+ 8 (t-2W) (1 + t/W) T(t,0,0) 
- (10t + 8W) B(0,t)^2
\nonumber \\ &&
- (36 t/W + 56 + 16 W/t) A(t) B(0,t)
+ (30 t^2/W + 42 t - 12W) B(0,t)
\nonumber \\ &&
- (40/W + 24/t) A(t)^2 
+ (30t/W + 84) A(t) 
- 39 W + 17 t/2
\nonumber \\ &&
- (n_Q-1)  W \bigl\{31 + 12 B(0,0) + 8 W M(0,0,0,0,0) \bigr \}
\biggr ] .
\label{eq:Delta2M2WQCD}
\eeq
The remaining, non-QCD, 2-loop contributions, are much more complicated,
involving a large number of terms. The form of the result 
is\footnote{Of the 78 coefficients $c_{j,k}^{(1,1)}$ 
for products of
1-loop integrals, 42 vanish.}
\beq
\Delta^{(2),{\rm non-QCD}}_W &=& 
\sum_i c^{(2)}_i I_i^{(2)}
+ \sum_{j\leq k} c_{j,k}^{(1,1)} I_j^{(1)} I_k^{(1)}
+ \sum_j c^{(1)}_j I_j^{(1)}
+ c^{(0)} .
\label{eq:Delta2M2WnonQCD}
\eeq
The coefficients $c^{(2)}_i$ and $c_{j,k}^{(1,1)}$ and $c^{(1)}_j$
and $c^{(0)}$ are given in electronic form in an ancillary file 
{\tt coefficients.txt} provided with the arXiv source for this article. 
These coefficients are written exclusively in terms of the quantities 
$W, Z, t, h, v^2$ [by using eqs.~(\ref{eq:defW})-(\ref{eq:defh})
to eliminate $g, g', y_t$, and $\lambda$], 
as well as the fixed parameters $N_c$, $n_Q$, and $n_L$.
The latter can each be set equal to 
3 in the Standard Model, but are kept general 
for checking purposes, and to tag the fermion loop contributions. 

It should be noted that the coefficients in the expression 
of the pole mass in terms of the 
basis integrals are not unique. This is because 
different basis integrals are related by special 
identities that hold when the squared mass arguments 
are not generic. These identities include
eqs.~(A.15)-(A.21) of ref.~\cite{Martin:2014cxa}, and
eqs.~(A.14), (A.15), and (A.17)-(A.20) in ref.~\cite{Martin:2003it}.

When setting $s \rightarrow W$ in eq.~(\ref{eq:M2Wpole}), one encounters singular
behavior in individual terms, associated with photon lines attached to a $W$ boson
propagator. In general, such potentially singular terms 
should cancel in the complex pole mass
\cite{Passera:1998uj}. They are dealt with here by using expansions such 
as\footnote{Eq.~(\ref{eq:B0W}) 
is used to eliminate $B(0,W)$ everywhere, explaining its absence in eqs.~(\ref{eq:onelooplist}) and (\ref{eq:Delta1loop}).}
\beq
B(0,W) &=& 1 - A(W)/W + (s-W) [1 + A(W)/W - \lnbar(W-s)]/W
\nonumber \\ &&
+ (s-W)^2 [-1 - A(W)/W + \lnbar(W-s)]/W^2 +
{\cal{O}}(s-W)^3
\label{eq:B0W}
\eeq
with $\lnbar(x) \equiv \ln(x/Q^2)$. Similar expansions of 2-loop basis functions
that have thresholds or pseudo-thresholds 
at $s=W$ are carried out using the differential equations listed in section IV of
ref.~\cite{Martin:2003qz}, using methods similar to those found in \cite{Caffo:1999nk}.
After doing so, all pole and logarithmic singularities
in $s-W$ that are found in individual Feynman diagrams cancel in the total eq.~(\ref{eq:M2Wpole}), an important check.

Several other helpful checks were performed on the calculation. 
First, single and double poles in $\epsilon$ cancel in 
$s^W_{\rm pole}$. This cancellation relies on 
agreement between the counter-terms 
$c^X_{\ell,n}$ (for $X = v, g, g', \lambda, y_t, g_3$) as extracted 
from the $\beta$ functions and
Higgs scalar anomalous dimension in the 
literature, and the coefficients of 
divergent parts of the loop integrations performed here. 
Second, I checked that logarithms of $G = m^2 + \lambda^2 v^2$ cancel. 
This is required for 
the absence of spurious imaginary parts that could occur when the 
renormalization scale is chosen so that $G<0$, and spurious
divergences that could occur for $G=0$. 
Third, I checked the absence of spurious imaginary parts of $s^W_{\rm pole}$;
note that $\Gamma_W$ must be identically 0 in the case $n_Q = 1$, 
$n_L = 0$, because in the Standard Model
the $W$ boson can only decay to 
lighter fermion doublets. This checks
cancellations between diagrams with Goldstone boson propagators
and the corresponding
Landau gauge vector propagator parts with poles at 0 squared mass.
Fourth, I checked that in each of the formal\footnote{None of these limits
are close to being realized in the real world.} limits that the quantities 
$W$, $Z$, $t$, $h$, $4W-h$, $4Z-h$, $4t-Z$, $t-W$, $t+W-Z$, or $t+W-h$ vanish, 
the whole expression for $s^W_{\rm pole}$ 
is finite and well-behaved, even though many of
the individual 2-loop coefficients in 
eqs.~(\ref{eq:Delta1loop}), (\ref{eq:Delta2M2WQCD}), and (\ref{eq:Delta2M2WnonQCD})
are singular in one or more of those limits.
This again reflects non-trivial relations between 
different basis integrals when squared mass arguments are not generic.
Finally, the result for 
$s^W_{\rm pole}$ was analytically
checked to be renormalization group invariant 
through terms of 2-loop order. In principle, this should be 
equivalent to the check of cancellation of $1/\epsilon$ poles, 
but in practice it tests many intermediate steps of the calculation. 
This check is written as:
\beq
0 \> = \> Q\frac{d}{dQ} s^W_{\rm pole} &=& 
\left [ Q \frac{\partial}{\partial Q}
- \gamma_\phi v \frac{\partial}{\partial v}
+ \sum_X \beta_{X} \frac{\partial}{\partial X} 
\right ] s^W_{\rm pole} ,
\label{eq:RGinvariance}
\eeq
where $X = \{g, g', \lambda, y_t, g_3\}$, and $\gamma_\phi$ is the anomalous
dimension of the Higgs field. It uses the derivatives of 
basis integrals with respect to the implicit argument $Q$, given in 
eqs.~(4.7)-(4.13) of ref.~\cite{Martin:2003qz}, and 
derivatives of the 1-loop basis integrals
with respect to squared mass arguments, given for example in
eqs.~(A.5) and (A.6) of ref.~\cite{Martin:2014cxa}. 
It also uses the beta functions and scalar anomalous dimension 
given in refs.~\cite{MVI,MVII,Jack:1984vj,MVIII}, \cite{FJJ}.
A corresponding numerical check of renormalization scale invariance is performed in
the next section.

\section{Numerical results\label{sec:num}}
\setcounter{equation}{0}
\setcounter{figure}{0}
\setcounter{table}{0}
\setcounter{footnote}{1}

The numerical computation of $s^W_{\rm pole}$ given by 
eqs.~(\ref{eq:M2Wpole})-(\ref{eq:Delta2M2WnonQCD}) is accomplished 
using the program {\tt TSIL}
\cite{TSIL}. This requires only 13 calls of the function \verb|TSIL_Evaluate| (which uses
Runge-Kutta solution of coupled differential equations 
to obtain multiple basis integral functions 
simultaneously) as well as relatively fast evaluations of the integrals for
which analytic formulas in terms of polylogarithms are known and incorporated 
in {\tt TSIL}. 

For purposes of illustration, consider a benchmark set of input data: 
\beq
v(M_t) &=& \mbox{246.647 GeV},
  \label{eq:inputvev}
\\
g(M_t) &=& 0.647550,
  \label{eq:inputg}
\\
g'(M_t) &=& 0.358521,
  \label{eq:inputgp}
\\
\lambda(M_t) &=& 0.12597,
  \label{eq:inputlambda}
\\
y_t(M_t) &=& 0.93690,
  \label{eq:inputyt}
\\
g_3(M_t) &=& 1.1666,
  \label{eq:inputg3}
\eeq
where $Q = M_t = 173.34$ GeV is the input renormalization scale.
The top Yukawa coupling and strong coupling constant were
taken from ref.~\cite{Buttazzo:2013uya} version 4, 
and the electroweak gauge couplings were taken from 
ref.~\cite{Degrassi:2014sxa}.
The VEV $v(M_t)$, which should minimize the radiatively corrected effective potential
in the scheme used here,
has also been chosen to approximately reproduce the experimental value
of the $Z$ boson physical mass, 
using a separate calculation (similar to the present one, and also in the pure
$\MSbar$ scheme) that I plan to report on soon. 
The Higgs self-coupling $\lambda$ was simultaneously chosen so as to 
also obtain a Higgs pole mass of
$M_h = 125.09$ GeV,
using the calculation of \cite{Martin:2014cxa} as implemented in
the program ${\tt SMH}$ \cite{SMHwebpages}, 
at an optimal renormalization scale $Q = 160$ GeV.
However, in the absence of a true global fit to available data,
it is important to emphasize that
the benchmark parameters chosen here should be viewed as illustrative,
rather than as a prediction of $M_W$.

\begin{figure}[!p]
\includegraphics[width=0.6\linewidth,angle=0]{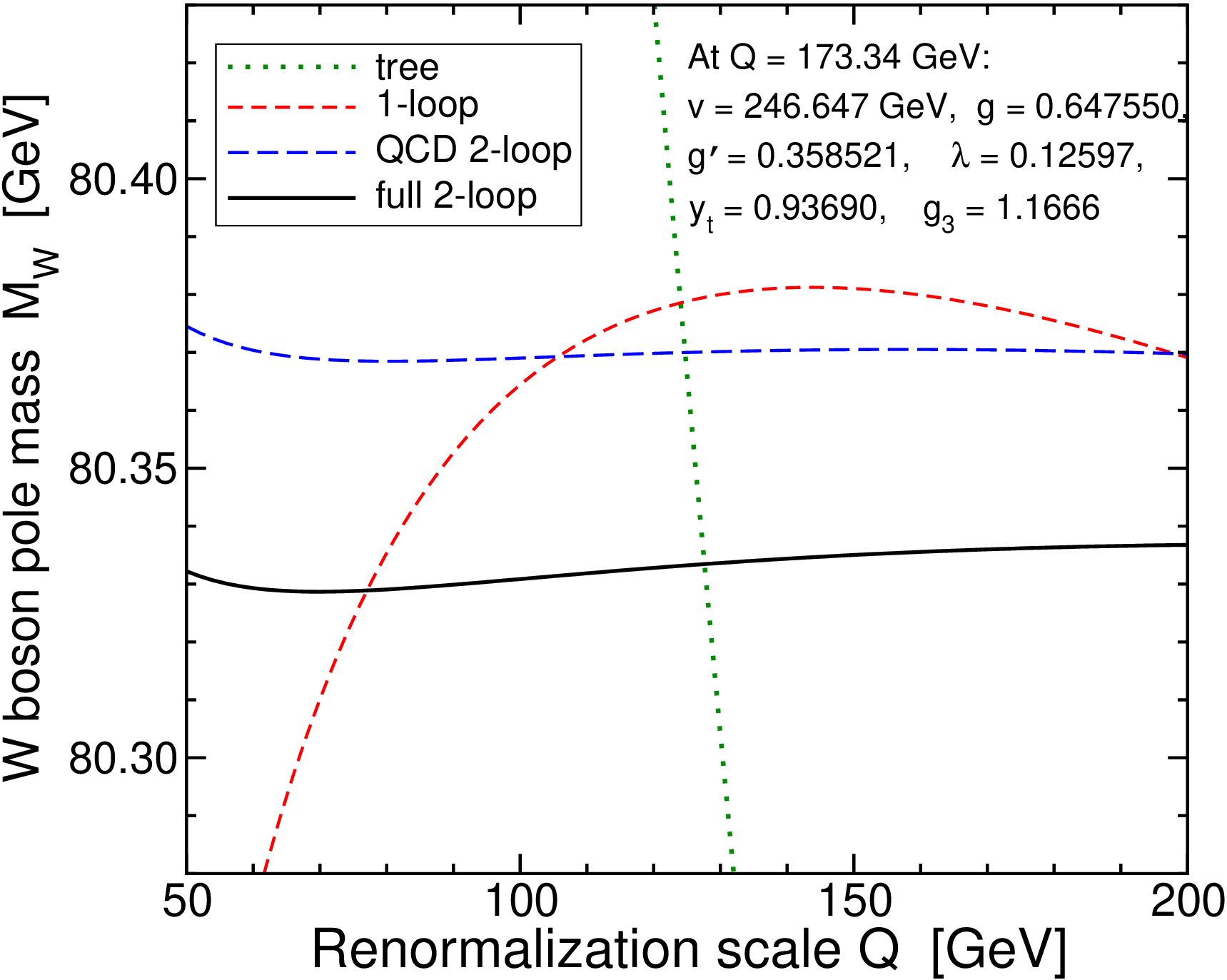}
\begin{minipage}[]{0.95\linewidth}
\caption{The mass $M_W$ of the $W$ boson, obtained from the
complex pole squared mass $s^W_{\rm pole} = M_W^2 - i \Gamma_W M_W $,
as a function of the renormalization scale $Q$ at which
$s^W_{\rm pole}$ is computed, in various approximations. 
The (green) dotted line is the tree-level result $W$, 
the (red) short-dashed line is the 1-loop result, the (blue) long-dashed line
is the result from the 1-loop and 2-loop QCD contribution, and the (black)
solid line is the full 2-loop order result. 
The input parameters $v, g, g', \lambda, y_t, g_3$ are 
obtained at the scale $Q$ by 3-loop renormalization group running, 
starting from eqs.~(\ref{eq:inputvev})-(\ref{eq:inputg3}).
Note that the usual Breit-Wigner mass $M_{W,{\rm exp}}$ is about 27 MeV larger
than $M_W$.
\label{fig:MWQ}}
\end{minipage}
\end{figure}

\begin{figure}[!p]
\includegraphics[width=0.6\linewidth,angle=0]{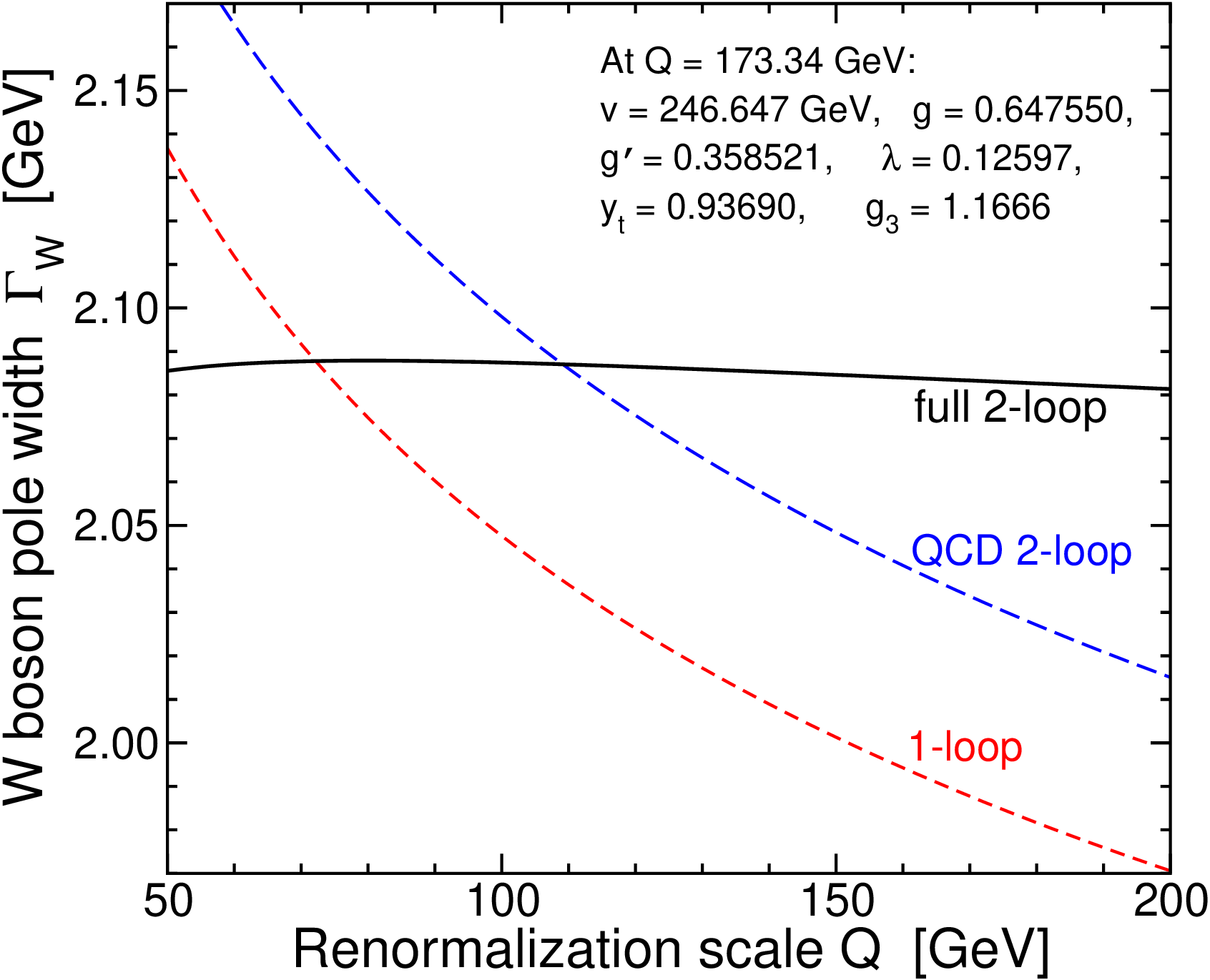}
\begin{minipage}[]{0.95\linewidth}
\caption{The width $\Gamma_W$ of the $W$ boson, obtained
from the
complex pole squared mass $s^W_{\rm pole} = M_W^2 - i \Gamma_W M_W $,
as in Figure \ref{fig:MWQ}.
The (red) short-dashed line is the 1-loop result, 
the (blue) long-dashed line
is the result from the 1-loop and 2-loop QCD contribution, and the (black)
solid line is the full 2-loop order result. 
\label{fig:GammaWQ}}
\end{minipage}
\end{figure}

The results for the renormalization scale dependences of $M_W$ and $\Gamma_W$
obtained 
from $s^W_{\rm pole} = M_W^2 - i \Gamma_W M_W $, in various approximations,
are shown in Figures \ref{fig:MWQ} and \ref{fig:GammaWQ}. 
To make the graphs, the input parameters 
$v,g,g',\lambda,y_t,g_3$ are run, using 3-loop beta 
functions \cite{Chetyrkin:2013wya,Bednyakov:2013eba},
from the input scale $M_t$ to the scale $Q$ on the horizontal axis, 
and $s^W_{\rm pole}$ is re-computed
at that scale. In the idealized case, $M_W$ and $\Gamma_W$ would be independent of
$Q$ if computed to sufficiently high order in perturbation theory.

In Figure \ref{fig:MWQ}, the (green) dotted line is the tree-level result $W$,
which shows a severe scale dependence, due to the running of $g$ and $v$. This
is still large, but reduced, in the 1-loop result, given by the (red) short-dashed line.
The majority of the remaining scale dependence is eliminated 
by including the QCD part of the 2-loop result from eq.~(\ref{eq:Delta2M2WQCD})
as shown in the (blue) long-dashed line. 
The (black) solid line shows the full 2-loop result. Note that despite the large
scale dependence of the 2-loop QCD correction, it is actually smaller than the 
2-loop non-QCD correction in magnitude except for $Q \lsim 85$ GeV, where the effect of
$\lnbar(t)$ starts to become large. 
The 2-loop non-QCD correction is of order 40 MeV, but is seen to have a quite mild scale dependence.

In Figure \ref{fig:GammaWQ} the (red) short-dashed line shows the
running of $\Gamma_W$ computed at 1-loop order. Adding in the 2-loop QCD contribution,
as shown by the (blue) long-dashed line, is a significant effect, but does not eliminate
the scale dependence, which is mostly due to the 
electroweak 1-loop renormalization group running of $g$ and $v$. However, including 
the 2-loop non-QCD corrections to $s^W_{\rm pole}$ greatly 
ameliorates the scale dependence, as it captures and compensates for 
most of the effect of running of $g$ and $v$. 

\begin{figure}[!t]
\includegraphics[width=0.6\linewidth,angle=0]{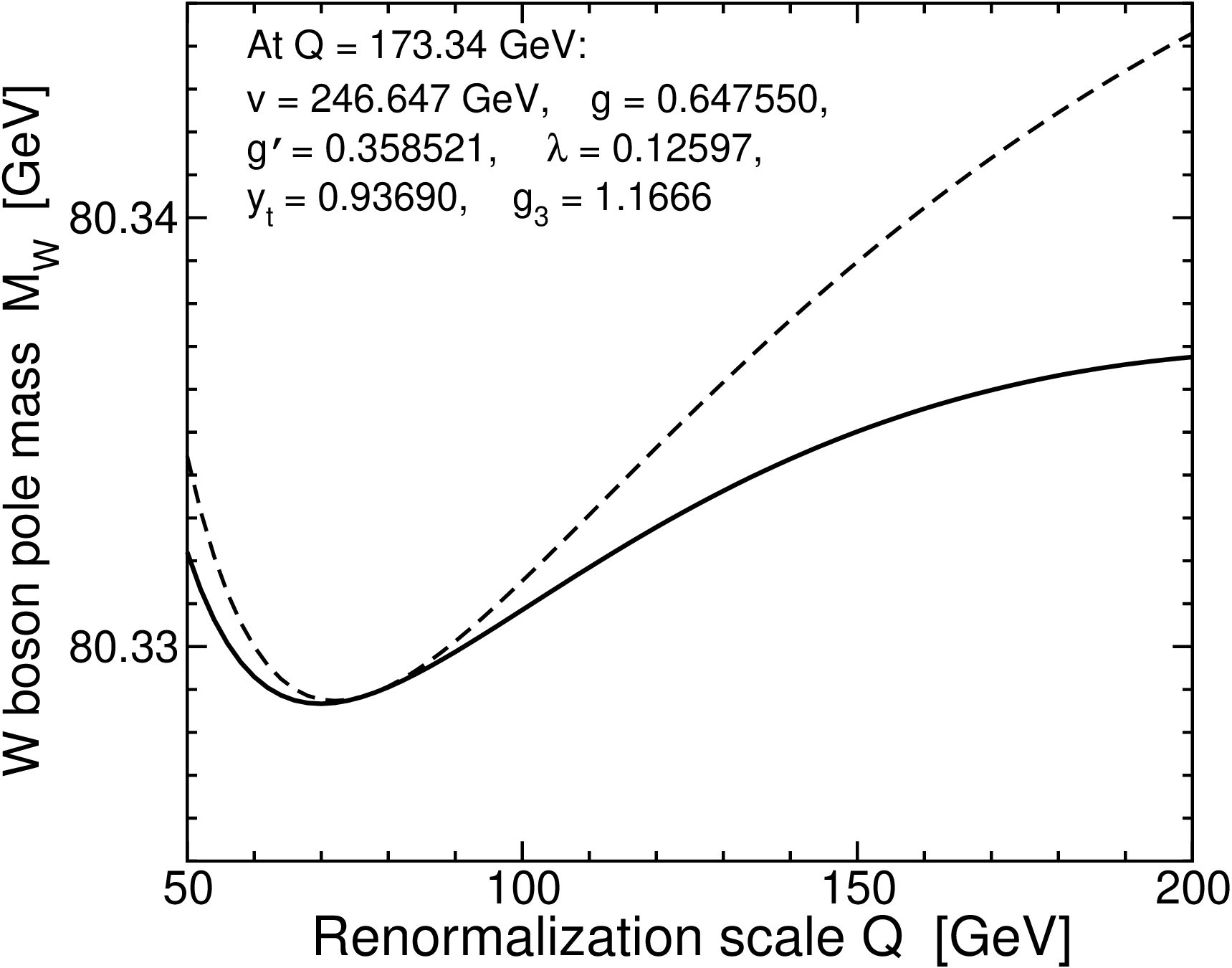}
\begin{minipage}[]{0.95\linewidth}
\caption{Close-up of the scale dependence of the mass $M_W$ of the $W$ boson, obtained
from the
complex pole squared mass $s^W_{\rm pole} = M_W^2 - i \Gamma_W M_W $,
as in Figure \ref{fig:MWQ}.
The solid line is the full 2-loop order result, while the
dashed line is the same, but after expanding the $\MSbar$ mass $t$ (in the 1-loop part
only) about $T = (173.34$ GeV$)^2$ to first order, using eq.~(\ref{eq:expT}). 
\label{fig:MWQexp}}
\end{minipage}
\end{figure}

For the range 50 GeV $<Q<$ 200 GeV, the deviations of $M_W$ and $\Gamma_W$ from 
their median values are both about $\pm 4$ MeV. 
For $M_W$, this is shown in close-up as the solid line in Figure \ref{fig:MWQexp}.
While this gives some lower bound on the remaining theory error 
(not counting the parametric errors in the inputs $v,g,g',\lambda,y_t,g_3$), 
it is always questionable to assume a direct relationship between scale dependence
and theory error. For another handle on the theory error, consider the following exercise.
In the top/bottom 1-loop contribution, the running top mass $t$ is used in propagators
in the pure $\MSbar$ scheme. However, once the result has been obtained, one can
expand $t$ about any other value, for example the top-quark pole mass $T$.
Doing so for the 1-loop contribution only is sensible, since $t$
only appears in propagators, not vertex couplings, 
in the 1-loop order $W$ boson self-energy. The relevant expansion is: 
\beq
f(0,t,W) = f(0,T,W) + (t-T)\bigl [
A(T) - 2 W + (T+W) B(0,T)
\bigr ]/2W + {\cal O}(t-T)^2. \phantom{xx}
\label{eq:expT}
\eeq
If this expansion is extended to, say, 4th order in $t-T$, 
then the results are easily checked to be nearly
indistinguishable from the original $f(0,t,W)$ without expansion. However, terminating
the expansion at linear order in $t-T$, as in eq.~(\ref{eq:expT}), can be considered
an alternative consistent 2-loop order result, if $t-T$ is treated as formally of 
1-loop order. 
This version of $M_W$ is shown as the dashed line in 
Figure~\ref{fig:MWQexp}. It clearly has a worse scale dependence, particularly at larger
$Q$, where $T-t$ becomes large. This suggests that the $\pm 4$ MeV scale dependence of the original (solid line)
pure $\MSbar$ calculation may be at least partly a fortunate accident.
The two curves agree near $Q =77$ GeV, where
the running top-quark mass $t$ equals the physical mass $T$. 

\section{Outlook\label{sec:outlook}}
\setcounter{equation}{0}
\setcounter{figure}{0}
\setcounter{table}{0}
\setcounter{footnote}{1}
\vspace{0.25cm}

In this paper I have reported the results for the complex pole mass of the $W$ boson
in the Standard Model in the pure $\MSbar$ scheme, with the vacuum expectation value,
defined as the minimum of the Landau gauge effective potential, taken as one of 
the input parameters. The organization of input and output parameters 
is quite different from previous works that use the on-shell scheme or hybrid 
$\MSbar$/on-shell schemes. The state-of-the art computations in these schemes,
see respectively e.g.~\cite{Awramik:2003rn} and \cite{Degrassi:2014sxa} 
and references therein, probably both attain a better theory error 
than the pure $\MSbar$ scheme, for now.
Moreover, a direct comparison of numerical results 
will need at least the corresponding 
results for the $Z$ boson, which I hope to report on soon. Both results will then be incorporated into a publicly available computer code together with the Higgs boson
mass code from \cite{Martin:2014cxa,SMHwebpages}.

Refs.~\cite{Jegerlehner:2001fb,Jegerlehner:2002em} and the very recent 
ref.~\cite{Kniehl:2015nwa} (which appeared as the present paper was being finished) 
also used the pure $\MSbar$ scheme to compute the complex pole mass of the $W$ boson.
However, attempts at direct comparison are complicated\footnote{Also, refs.~\cite{Jegerlehner:2001fb,Jegerlehner:2002em} use expansions in 
$1/4 - \sin^2\theta_W$ and $Z/h$ and $Z/t$ 
(in the notation of the present paper), which
further increases the difficulty in making a direct comparison.}  
by the fact that these papers used a different definition of the VEV, namely
$v_{\rm tree}^2 = -m^2/\lambda$, rather than 
$v$ that minimizes the full radiatively corrected effective potential
as made here 
(and, for example, refs.~\cite{Martin:2014cxa} 
and\footnote{However,
  Ref.~\cite{Degrassi:2014sxa} uses Feynman gauge instead of Landau gauge, so the VEV 
  referred to in that paper will also not be the same thing as $v$ in the present paper.
  Note that using $v$ requires choosing a gauge-fixing prescription; choosing 
  Landau gauge 
  has the advantage that the effective potential is much simpler.}
\cite{Degrassi:2014sxa}). 
The choice of using $v_{\rm tree}$ requires including non-trivial tadpole diagrams,
unlike the choice of expanding around $v$ where 
the sum of Higgs tadpole diagrams (including the tree-level tadpole) simply vanishes. 
This means that already at 1-loop order, the expressions appear different. 
Compared to $\Delta^{(1)}_W/(16 \pi^2)$ in eq.~(\ref{eq:Delta1loop}) of the present paper,
the sum of the bosonic contributions in
eq.~(B.2) of ref.~\cite{Jegerlehner:2001fb} and the fermionic contributions in (B.2) of
ref.~\cite{Jegerlehner:2002em} differs by:
\beq
\frac{g^2}{16 \pi^2 h} \left [ -2 N_c t A(t) + 
\frac{3}{4} h A(h) + 3 W A(W) + 2 W^2 + \frac{3}{2} Z A(Z) + Z^2 \right ] .
\label{eq:diff1loop}
\eeq
This is simply because the tree-level terms are also different, namely $g^2 v^2/4$ in the present paper and $g^2 v^2_{\rm tree}/4$ 
in refs.~\cite{Jegerlehner:2001fb,Jegerlehner:2002em,Kniehl:2015nwa}.
To 1-loop order accuracy, the two expressions for the pole mass
can easily be checked to be the same,
by using eq.~(\ref{eq:resummedmincon}) above, 
but establishing the connection at 2-loop order 
would require a somewhat non-trivial re-expansion using the 2-loop relation between 
$v_{\rm tree}^2$ and $v^2$.

Note that, in general, expanding around $v_{\rm tree}$ rather than $v$ 
has the effect of making the perturbative
expansion parameter be $\displaystyle\frac{N_c y_t^4}{16 \pi^2 \lambda}$, 
rather than the usual
$\displaystyle\frac{N_c y_t^2}{16 \pi^2}$, for the terms leading in the top mass.
This can be seen in the presence of the first term 
in eq.~(\ref{eq:diff1loop}); in contrast, there is no 
$g^2 t A(t)/h$ term in eq.~(\ref{eq:Delta1loop}). As mentioned above as one of the checks,
at two-loop order there is also no behavior like $t^3/h^2$ or $t^2/h$ (or any other pole
singularity in $h$, or $W$ or $Z$) in 
$\Delta^{(2),{\rm non-QCD}}_W$ in eq.~(\ref{eq:Delta2M2WnonQCD}).
As another example, see the discussion surrounding eqs.~(4.34)-(4.40)
in ref.~\cite{Martin:2014bca}, 
where the terms of order 
$\displaystyle\left (\frac{y_t^4}{16 \pi^2 \lambda} \right)^\ell$
in the relation between $v_{\rm tree}$ and $v$ 
are explicitly identified for loop orders $\ell = 1,2,3$ in the limit 
$y_t^2 \gg \lambda$ in the case $g = g' = 0$.
Not surprisingly, expanding around the radiatively corrected VEV 
leads to faster convergence than expanding around the tree-level VEV, 
at least formally, although both expansions
should converge given enough loop orders, since  
$\displaystyle\frac{N_c y_t^4}{16 \pi^2 \lambda}$ is
still numerically small.

\vspace{0.2cm}

It would clearly be useful to include the 3-loop contributions to 
$W$ and $Z$ complex pole masses
in the pure $\MSbar$ scheme, so that theory errors can be made unambiguously 
much smaller 
than all relevant experimental errors. 
Here it should be remarked that it is not at all obvious
that the parametrically QCD-enhanced 
contributions at 3-loop order will be the largest, especially considering
that this was not the case at 2-loop order. 
A possible scenario is that the QCD-enhanced
contributions will have the largest renormalization scale dependence, but not the largest
magnitude, since this is what happened at 2-loop order. It seems feasible to eventually
include all 3-loop contributions to $s^W_{\rm pole}$ in the pure $\MSbar$ scheme, 
although to do so without using mass expansions or approximations may require developing new methods for treating 3-loop self-energy contributions.

\vspace{0.2cm}

\noindent {\it Acknowledgments:} 
This work was supported in part by the National
Science Foundation grant number PHY-1417028. 


\end{document}